\documentclass[12pt,preprint]{aastex}
\begin{document}

\title{Ionized Outflows from Compact Steep Spectrum Sources}
\author{Hsin-Yi Shih\altaffilmark{1}, Alan Stockton\altaffilmark{1},Lisa Kewley\altaffilmark{1,2}}
\altaffiltext{1}{Institute for Astronomy, University of Hawai'i, 2680 Woodlawn Dr, Honolulu, HI 96822}
\altaffiltext{2}{The Australian National University}
\email{hsshih@ifa.hawaii.edu, stockton@ifa.hawaii.edu, lisa.kewley@anu.edu.au}

\begin{abstract}

Massive outflows are known to exist, in the form of extended emission-line regions (EELRs), around about one-third of powerful FR II radio sources. We investigate the origin of these EELRs by studying the emission-line regions around compact-steep-spectrum (CSS) radio galaxies that are younger (10$^3$ to 10$^5$ years old) versions of the FR II radio galaxies. We have searched for and analyzed the emission-line regions around 11 CSS sources by taking integral field spectra using GMOS on Gemini North. We fit the  [\ion{O}{3}] $\lambda 5007$ line and present the velocity maps for each detected emission-line region. We find, in most cases, that the emission-line regions have multi-component velocity structures with different velocity dispersions and/or flux distributions for each component. The velocity gradients of the emission-line gas are mostly well aligned with the radio axis, suggesting a direct causal link between the outflowing gas and the radio jets. The complex velocity structure may be a result of different driving mechanisms related to the onset of the radio jets. We also present the results from the line-ratio diagnostics we used to analyze the ionization mechanism of the extended gas, which supports the scenario where the emission-line regions are ionized by a combination of AGN radiation and shock excitation. 

{\it Subject Keywords:} galaxies: kinematics and dynamics
\end{abstract}

\maketitle

\section{Introduction}

Galactic feedback mechanisms are an integral part of galaxy evolution models. Feedback regulates the star formation rate and black hole growth, which may in turn affect many other physical properties of galaxies, such as bulge mass and stellar populations. In this paper we investigate one form of feedback possibly driven by the radio jets and attempt to answer the following questions: (1) Do we see any evidence of radio jets driving the feedback in the morphology and velocity fields of the outflowing gas and (2) What do the emission line diagnostics tell us about the ionization mechanism and the physical properties of the outflowing gas?

Extended emission-line regions (EELRs), typically AGN ionized gas clouds on the scales of few tens of kpc, are present around a significant fraction of powerful FRII radio sources \citep[e.g.,][]{stockton87}. They consist of massive clouds of ionized gas extending beyond the host galaxies. Over the years since their discovery, various formation mechanisms have been suggested for EELRs, including tidal debris, cooling flows, and outflows driven by starbursts or the AGN themselves. We discuss these possibilities briefly here; see, e.g., \citet{fu09a} for a more detailed discussion. 

Comparison of deep continuum images with the EELR distributions shows essentially no correlation between the ionized gas and the stars \citep[e.g.][]{stockton02}. This mismatch is not necessarily a problem for the tidal debris scenario, since the gas is collisional and the stars are not. But \citet{fabian87} pointed out that a real problem for the tidal interpretation was that the ionized gas clouds, if unconfined, would dissipate in a sound-crossing time of $\sim10^4$--$10^6$ years. They suggested instead that the $10^4$ K ionized gas had condensed as a cooling flow from a hot ($\sim10^7$ K) halo, which also served to confine the cooler gas. However, \citet{stockton06}, from Chandra X-ray observations of four EELR quasars, placed an upper limit to the pressure of any such halo gas well below that needed to confine the  EELR gas. 

A possible way to rescue the tidal origin hypothesis is to assume that most of the extended ionized gas comes from self-gravitating giant molecular clouds (GMCs), where the gas is ablated from the surface of the clouds by the ionizing radiation. This an attractive suggestion, but there appear to be problems with this picture as well: (1) because of their high densities, the GMCs would tend to act like collisionless particles and would therefore follow the stellar distribution, which, as mentioned above, is the not generally the case for EELRs; and (2) masses of luminous EELRs typically range from a few $10^9$ M$_{\odot}$ to a few $10^{10}$ M$_{\odot}$, which would require at least $\sim1000$ GMCs with masses similar to those found in our Galaxy.

Most EELRs show some clouds with radial velocities $>400$ km s$^{-1}$ relative to the systemic velocity \citep[e.g.][]{fu09a}, which are higher than can be reasonably explained by gravitational kinematics. These high-velocity clouds suggest an outflow origin for EELRs. However, since the star formation rate in most of the EELR quasar host galaxies is too low to drive superwinds that could produce EELRs \citep{fu08,fu09b}, the most plausible source for energy input is the AGN itself. In particular, since luminous ($L_{Extended[O III]} > 5\times10^{41}$ erg s$^{-1}$) EELRs are found almost exclusively around FR II radio sources \citep{boroson85,stockton87}, which have high speed radio jets, it is likely that powerful radio jets are the dominant driving mechanism behind the outflows. In such an outflow, turbulent shocks can produce transient dense ($\sim500$ cm$^{-3}$) filaments in an otherwise low-density ($\sim1$ cm$^{-3}$) gas, eliminating the need for confinement of the dense clouds \citep[e.g.][]{stockton02,fu09a}.

To better understand the origin of the EELRs, we need to observe them in young radio sources where the outflow is still at its infancy. Compact steep-spectrum (CSS) sources are young \citep[$\sim 10^{3} - 10^{5}$ yr ][]{odea98, desilva99} radio sources where the radio structure does not extend beyond the optical scale of the galaxy. This study is partly motivated by the intriguing properties of the ionized outflow gas in the CSS quasar 3C\,48. This outflow was first noticed by \citet{cha99} and by \citet{can00}, and studied in more detail with GMOS IFU spectra by \citet{sto07}. While the peak emission from the outflow is located near the base of the radio jet, it is clearly extended over a wide solid angle and is even seen in absorption in the spectrum of the quasar itself \citep{gup05}. While a potential concern is that, with a quasar, projection effects might magnify the angular extent of the outflow, \citet{wil91} have pointed out that the weakness of the central radio source in 3C\,48 argues strongly that the radio axis is not close to the line of sight. In addition, the velocity width of the resolved emission is $>1500$ km s$^{-1}$, indicating gas flows over a large solid angle. Nevertheless, 3C\,48 is by no means a typical quasar \citep[e.g.,][]{sto07}, and there remains the possibility that its outflow is not representative of that of other CSS sources. (Note: The emission-line regions around CSS sources extend from 4 to 12 kpc radially from the nucleus. Their spatial extent partially overlaps with the sizes of the classical narrow-line regions ($\sim 5$ kpc), and they do not reach the tens of kpc scale of the classical EELRs around more evolved radio sources. Nevertheless, for convenience, the term ``EELR" in the rest of this paper refer to the smaller extended emission-line regions found around CSS sources, unless they are specified as classical EELRs.)

EELRs are present around a large fraction of CSS sources \citep[e.g.,][]{axon00}. Some of these sources have been studied with long-slit spectra \citep[e.g.,][]{holt08,holt09}. At least part of the EELR was shown to be a high velocity outflow, while other emission could have resulted from ionizing the interstellar medium (ISM). However, long-slit spectra do not give enough information on the velocity structure to determine its angular variation with respect to the radio structure, nor do they give detailed information on the morphologies of the extended emission-line clouds. In this study, we go a step further and analyze two dimensional spectra of these objects obtained from integral-field spectroscopy. This not only gives us a more complete picture of the EELRs' velocity structure, but also gives us the morphology of the EELRs and, to some extent, that of the host galaxies.

\section{Observations and Data Reduction}

We observed 11 CSS radio galaxies with the Integral Field Unit (IFU) of the Gemini Multi-Object Spectrograph (GMOS) on Gemini North. Our sample is compiled from the literature, including all CSS radio galaxies with $0.2 < z < 0.8$ (except for 3C\,124, which is at $z=1.083$ and is included in our sample because it clearly has a resolved EELR in narrow-band HST images) accessible from Mauna Kea. The sources and observation parameters are shown in Table 1. Previous studies of EELRs around CSS sources mostly focused on quasars. We chose to focus on radio galaxies because their radio jets are believed to be, on average, more aligned with the plane of the sky, which makes studying the spatial correlation between the EELRs and the radio morphologies more straightforward. We search for extended [\ion{O}{3}] $\lambda 5007$ emission around each object. For 3C\,124, the [\ion{O}{3}] $\lambda 5007$ line is shifted out of the visible range and we observed the  [\ion{O}{2}] $\lambda 3727$ line instead. All sources have existing radio maps.

All targets were observed using the single-slit mode of the IFU, which has a field of view of $3\farcs5 \times 5\arcsec$. We used the B600 grating for objects with $z < 0.6$ and R400 grating for objects with $z > 0.6$. The gratings have resolutions of $R = 1688$ and $R = 1918$ respectively. The central wavelengths of each object are tuned so that the data covers the rest frame wavelength of 3140\AA $-$ 5400\AA. Each object has at least 2 hours on-target exposures and four dither positions ($1800$ s exposure time per position). The observations were planned in a way such that the [\ion{O}{3}] $\lambda \lambda 4959, 5007$ and H$\beta$ lines land well within the detectors, avoiding the chip gaps. The blueward sides of the spectra encompass the [\ion{O}{2}] $\lambda 3727$ line. All data were taken on clear, dark nights with seeing $< 0.8 \arcsec$. The total science field that was used for analysis is $4\arcsec \times 5\arcsec$. 

The data were reduced using the Gemini IRAF package following the standard steps: bias subtraction, flat-fielding, wavelength calibration, sky subtraction, flux calibration, and spectral extraction. The reduced 2-D images were assembled into a 3-D data cube with coordinates (x, y, $\lambda$) at the original resolution of 0\farcs2 per spatial pixel. The cubes at different dither positions were shifted and combined into one final data cube for each target. Some lower signal-to-noise pixels were binned to ensure that the lines we used for analysis are at roughly the same S/N.

\begin{deluxetable}{lcccccc}
\tablecaption{CSS Galaxy Sample}
\tablewidth{0pt}
\tablecolumns{7}
\tablehead{\colhead{Object Name} & \colhead{$z$} & \colhead{Exposure}  & \colhead{Emission?}  & \colhead{Extended?}& \colhead{Radio size} & \colhead{S$_{1.4GHz}$} \\
\colhead{} & \colhead{} & \colhead{Time} & \colhead{} & \colhead{} & \colhead{(arcsec)}  & \colhead{Jy}}
\startdata
3C\,49 & 0.621\phantom{00} (2) & $1800\times5$ & Yes & Yes & 1.0  (8) & 2.74\\
PKS\,0023$-$263 & 0.32162 (3) & $1800\times5$ & Yes & Yes & 1.0 (9) & 8.75\\
PKS\,2135$-$209 & 0.63634 (3) & $1800\times4$ & Yes & Yes & 0.3 (9) & 3.73\\
3C\,268.3 & 0.37171 (3) & $1800\times4$ & Yes & Yes & 2.0 (8) & 3.72\\
3C\,303.1 & 0.27040 (3) & $1800\times4$ & Yes & Yes & 1.9 (10)& 1.88\\
PKS\,1306$-$09 & 0.46685 (3) & $1800\times4$ & Yes & Yes & 0.4 (9) & 4.24\\
3C\,93.1 & 0.243\phantom{00} (6) & $1800\times4$ & Yes & Yes & 0.6  (8)& 2.36 \\
3C\,124 & 1.083\phantom{00} (7) & $1800\times4$ & Yes & Yes & 1.3 (11) & 1.14\\
4C\,46.25 & 0.7428\phantom{0} (5) & $1800\times4$ & Yes &  No & 1.2 (12)& 1.15\\
4C\,45.22 & 0.7621\phantom{0} (4) & $1800\times4$ & No & No & 1.1 (13)& 0.71 \\
B3\,1233+418 & 0.250\phantom{00} (1) & $1800\times4$ & Yes & No\ & 1.3 (13)& 0.69\\
\enddata
\tablerefs{Sources of redshifts: (1) \citet{mur99}; (2) \citet{spi85}; (3) \citet{holt08}; (4) \citet{tho94}; (5) \citet{max95}; (6) \citet{sar77}; (7) \citet{hew91}. Radio data references: (8) \citet{akujor91}; (9) \citet{tzioumis02}; (10) \citet{akujor95}; (11) \citet{privon08}; (12) \citet{dallacasa02}; (13) \citet{rossetti06}. Radio power reference: \citet{condon98}.}
\end{deluxetable}

\begin{deluxetable}{lccc}
\tablecaption{Radio and Optical Properties}
\tablewidth{0pt}
\tablecolumns{4}
\tablehead{\colhead{Object Name} & \colhead{Radio PA}  & \colhead{EELR PA} & \colhead{Velocity Gradient }\\
\colhead{} & \colhead{}  & \colhead{} & \colhead{Axis PA}}
\startdata
3C\,49 & $-88$ & $-96 \pm 20$ &\\
 \phantom{00}Broad 1& -- &  -- &$-83 \pm 20$ \\
 \phantom{00}Narrow 1&  --  & -- & $-21 \pm 10$ \\
 \phantom{00}Narrow 2 & --   & -- & $-109 \pm 13$ \\
 \phantom{00}Narrow 3 &  --  & -- & No clear gradient \\
 
PKS\,0023$-$263 & $-34$ & $-44 \pm 2$&\\
 \phantom{00}Broad 1& -- & -- & $-87 \pm 8$ \\
 \phantom{00}Narrow 1&  -- & -- & $-63 \pm 2$ \\
 
PKS\,2135$-$209   & $52$ & Circular &\\
 \phantom{00}Broad 1& -- & -- & $47 \pm 10$ \\

3C\,268.3  & $-19$ & $-37 \pm 5$ &\\
 \phantom{00}Broad 1& -- & -- & $-31 \pm 5$ \\
 \phantom{00}Narrow 1&  -- & -- & $-35 \pm 5$ \\

3C\,303.1 & $-50$  & $-32\pm 8$& \\
 \phantom{00}Broad 1& -- &  -- &$-51 \pm 5$ \\
 \phantom{00}Narrow 1&  -- & -- & $-48 \pm 9$ \\

PKS\,1306$-$09  & $-45$ & $-60 \pm 4$& \\
 \phantom{00}Narrow 1 \& 2& -- & -- & $-57 \pm 5$ \\

3C\,93.1 & Not well defined & $31 \pm 6$&\\
 \phantom{00}Broad 1& -- &  -- &$21 \pm 10$ \\
 \phantom{00}Narrow 1&  -- & -- & $22 \pm 5$ \\

3C\,124 & $7$ & $-1 \pm 5$&  \\
 \phantom{00}Narrow 1 \& 2& -- & -- & $4 \pm 13$ \\
\enddata
\tablerefs{Radio PAs measured from radio data cited in Table 1.}
\end{deluxetable}

\section{Data Analysis}

To properly analyze the emission-line regions, we first need to subtract the continuum contributions to the emission-line slices in the data cube. For each target, we take the median of $\sim 100$ slices to the blueward of the H$\beta$ line and the median of $\sim 100$ slices to the redward of the [\ion{O}{3}] $\lambda 5007$ line to produce a red and a blue continuum image. We then use a simple linear interpolation to estimate and subtract the continuum contribution from each slice of data containing line emission. Eight out of the eleven objects show clearly detected EELRs. 4C 46.25 and B3 1233+418 show faint emission lines,  but it is unclear whether they are extended. No line emission is detected for 4C 45.22. 

\begin{figure*}[t]
\centering
\includegraphics[width=6.in]{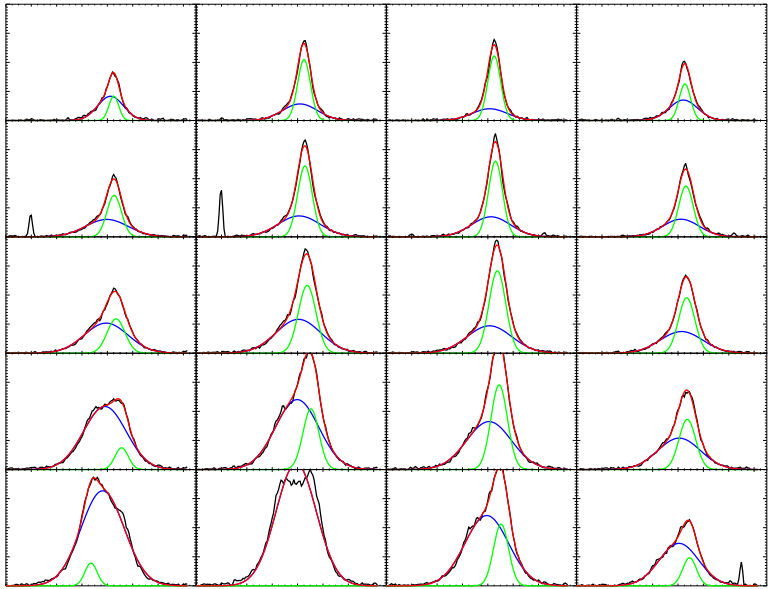}
\caption{Fitting the [\ion{O}{3}] $\lambda 5007$ profiles for each spaxel. One broad and one narrow velocity component is required to minimize residuals. The black line shows the data, the blue line shows the narrow component, the green line represent the broad component and the red line is the total fit. Only a small portion around the central region is shown. }
\label{3C303_1fits}
\end{figure*}

Before extracting information from the emission lines, we used the Voronoi binning method \citep{cappellari03} to bin the data, where necessary, to ensure that the [\ion{O}{3}]$ \lambda 5007$ line has a S/N of at least 50 in all spectra. We used MPFIT (Markwardt 2008) to extract kinematic information from the [\ion{O}{3}] $\lambda 5007$ line. We started with an one-component Gaussian fit for each spectrum and added more components when needed. Most objects needed two components for a decent fit. PKS 2135$-$209  is the only object that could be fitted with a single Gaussian component throughout. 3C 49, on the other hand, clearly has four peaks in the emission line profiles and needed four Gaussian components to minimize the residuals. The fits gave us information on the velocity and velocity dispersion of the ionized gas at different spatial positions. An example of a fit for the central portion of 3C 303.1 is found in Figure \ref{3C303_1fits}.

\section{Results}

\begin{figure*}[t]
\centering
\includegraphics[width=6.in]{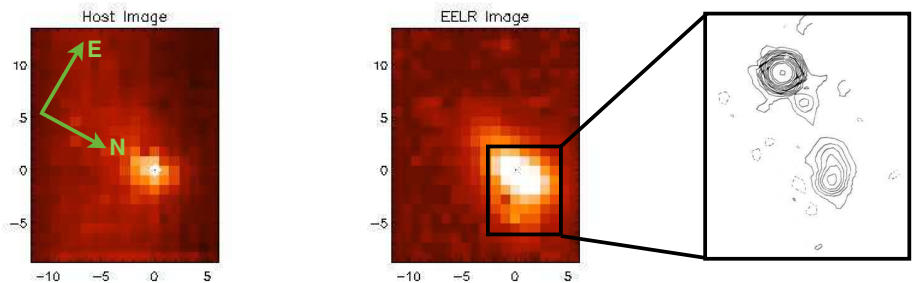}
\caption{{\bf 3C 49} Left - host galaxy image made from taking the median of continuum slices immediately blueward and redward of the [\ion{O}{3}] line. Middle: EELR image made by summing all the slices on the [\ion{O}{3}] $\lambda5007$ line. Right: Enlarged MERLIN 5 GHz  radio image from \citet{akujor91}. The observation of this object is taken at a PA of 240\arcdeg; north and east are labeled in the left panel. The numbers along the axes in this and subsequent images are distance from the continuum peak in kpc.}
\label{3C49map1}
\end{figure*}

\begin{figure*}[t]
\centering
\includegraphics[width=6.in]{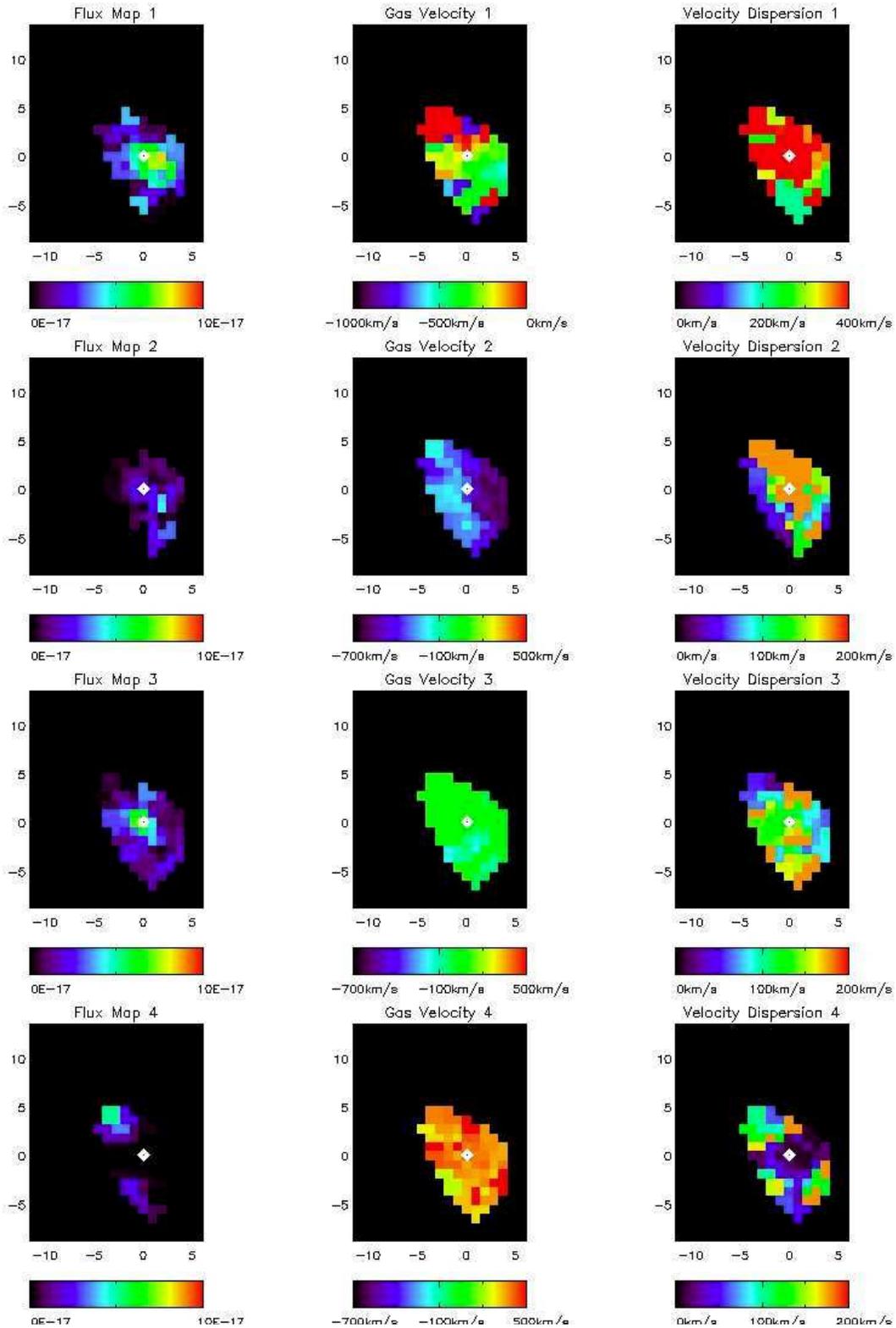}
\caption{{\bf 3C 49} Top row: left - flux map for the broad velocity component, the flux scale is in erg s$^{-1}$ cm$^{-2}$. Middle - velocity map of the broad velocity component, velocity is given in km s$^{-1}$. Right - velocity dispersion map of the broad velocity component, velocity dispersion is given in km s$^{-1}$. Next three rows: same as top row but for the three narrow velocity components. }
\label{3C49map2}
\end{figure*}

\begin{figure*}[t]
\centering
\includegraphics[width=6.in]{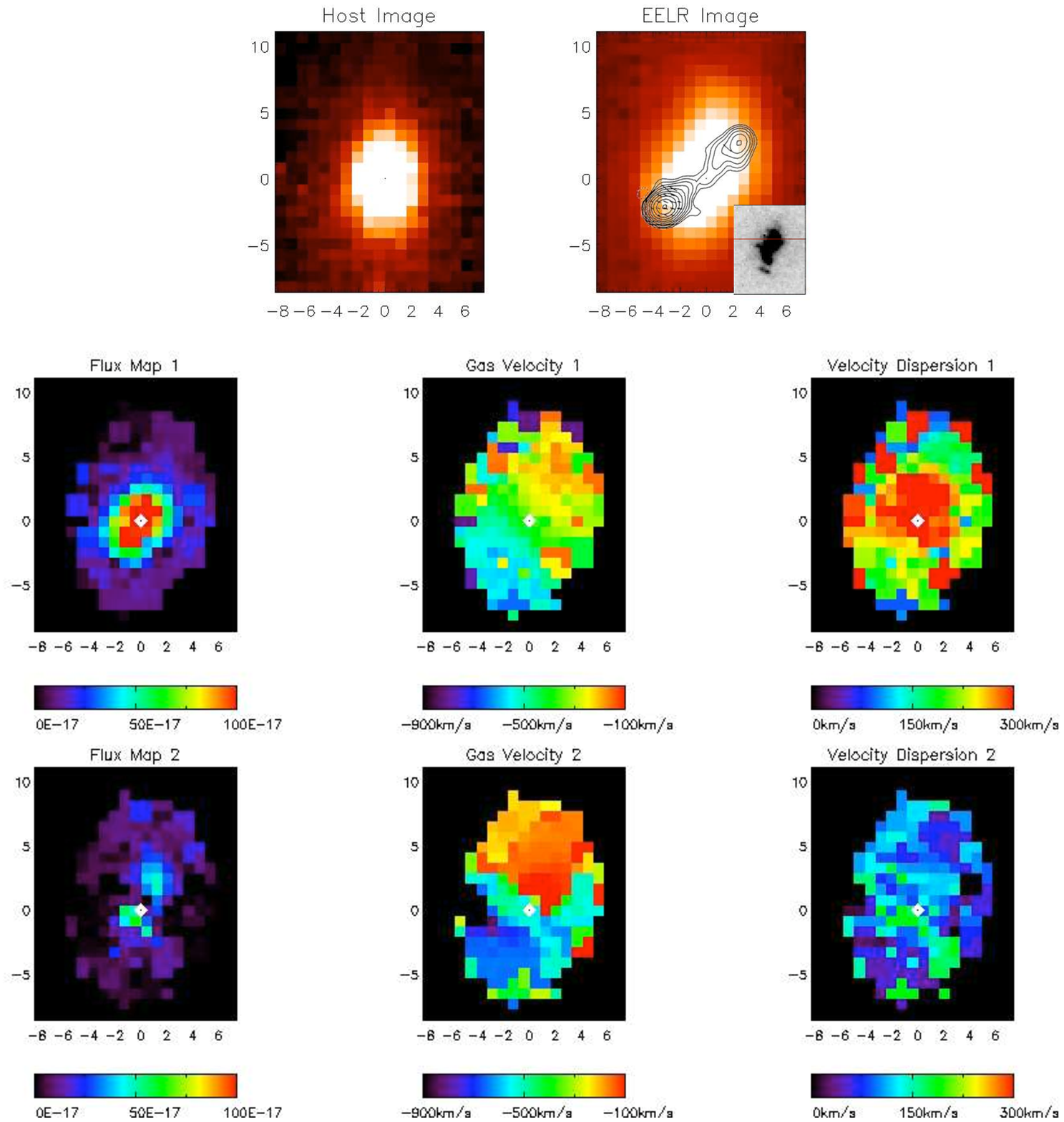}
\caption{{\bf 3C 303.1} Top row: Left - host galaxy image. Right - [\ion{O}{3}] image with MERLIN 8.4 GHz radio image from \citet{akujor95} overlaid. The inset on the right panel is the HST image published in \citet{axon00}, the physical size of the inset is the same as the FOV of our data cube. Middle row: left - flux map for the broad velocity component, middle - velocity map of the broad velocity component, right - velocity dispersion map of the broad velocity component. Bottom row: same as middle row but for the narrow velocity component. This image and all following images are oriented with north up and east to the left.}
\label{3C303_1map}
\end{figure*}

\begin{figure*}[t]
\centering
\includegraphics[width=6.in]{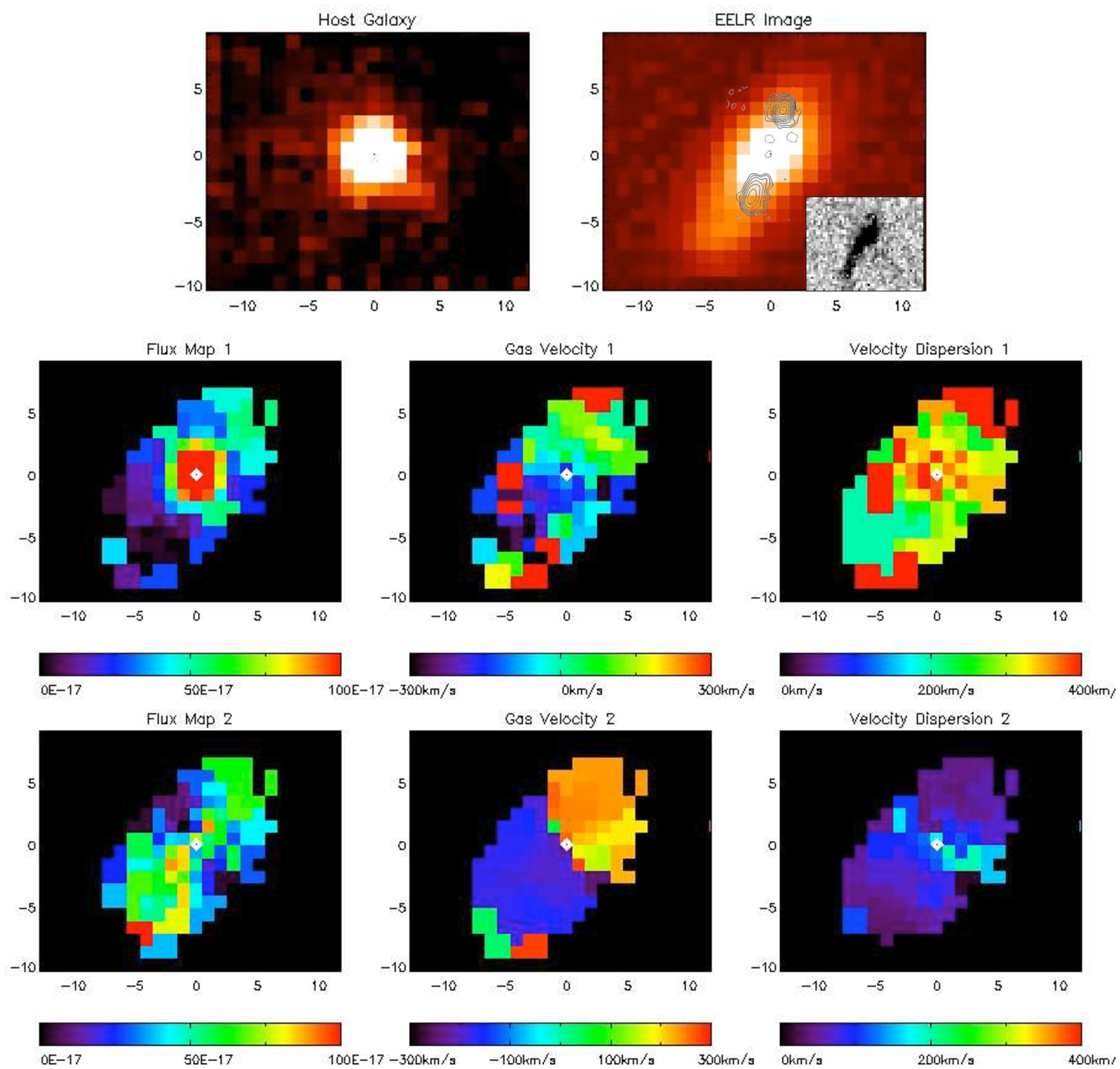}
\caption{{\bf 3C268.3} Top row: Left - host galaxy image. Right - [\ion{O}{3}] image with MERLIN 5 GHz  radio image from \citet{akujor91}. The inset on the right panel is the HST image published in \citet{axon00}, the physical size of the inset is the same as the FOV of our data cube. Middle row: left - flux map for the broad velocity component, middle - velocity map of the broad velocity component, right - velocity dispersion map of the broad velocity component. Bottom row: same as middle row but for the narrow velocity component. }
\label{3C268_3map}
\end{figure*}

\begin{figure*}[t]
\centering
\includegraphics[width=6.in]{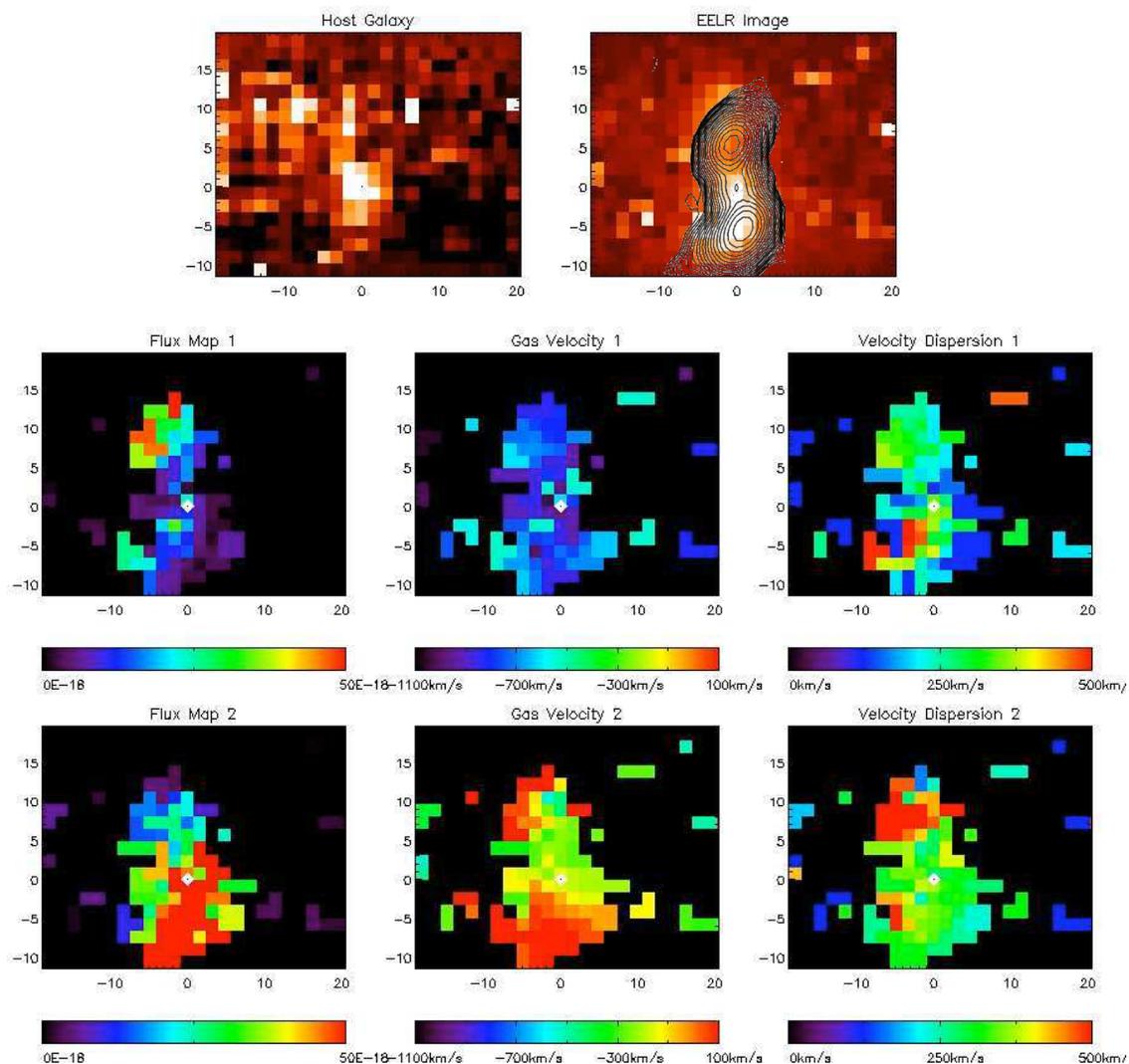}
\caption{{\bf 3C 124} Top row: Left - host galaxy image made from taking the median of continuum slices immediately blue-ward and red-ward of the [\ion{O}{2}] line. Right - EELR image for 3C 124 made by summing all the slices on the [\ion{O}{2}] $\lambda3727$ line, with VLA 4.885 GHz  radio image from \citet{privon08}. Middle row: left - flux map for the broad velocity component, middle - velocity map of the broad velocity component, right - velocity dispersion map of the broad velocity component. Bottom row: same as middle row but for the second velocity component. }
\label{3C124map}
\end{figure*}

\begin{figure*}[t]
\centering
\includegraphics[width=6.in]{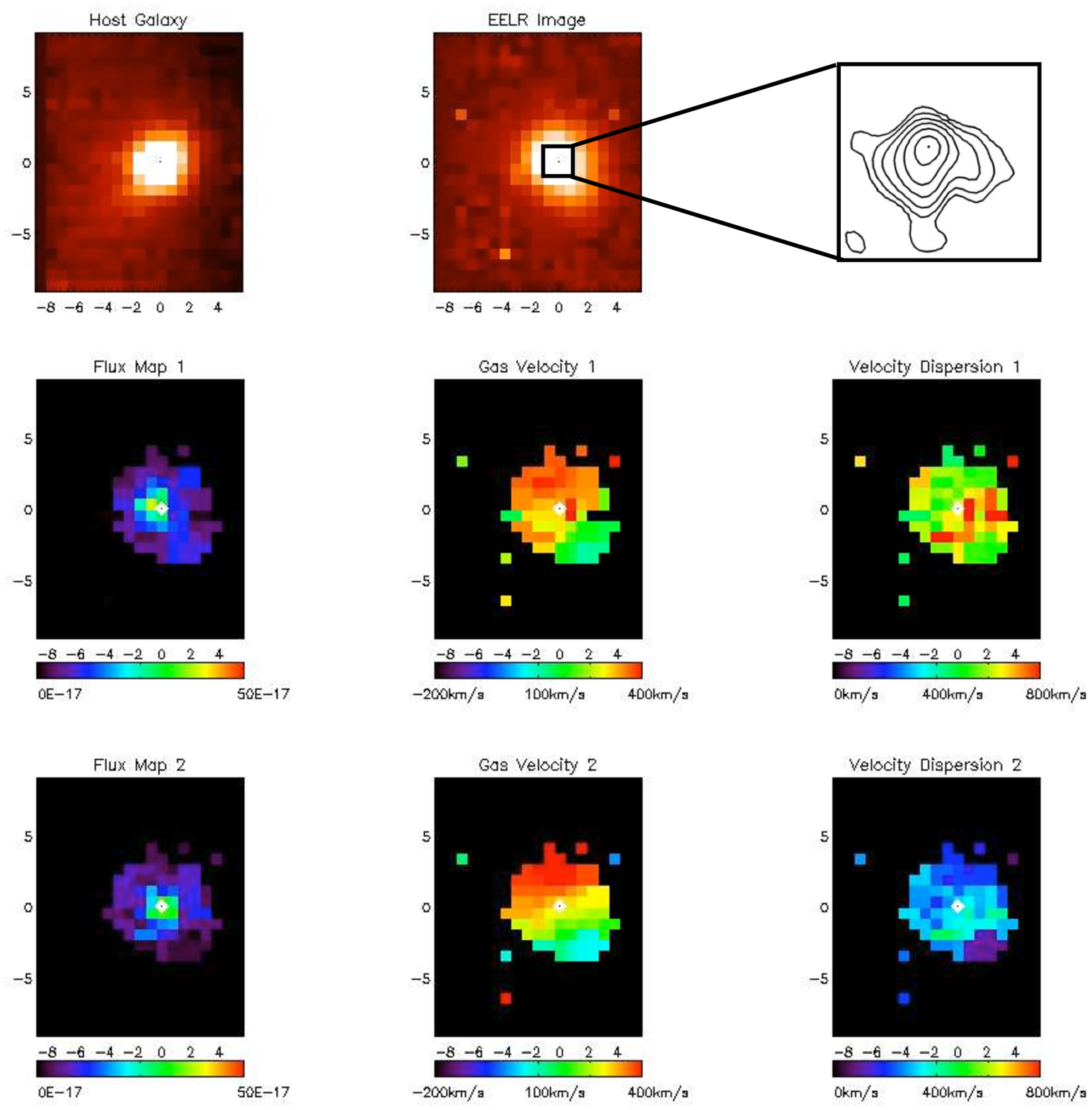}
\caption{{\bf 3C 93.1} Top row: Left - host galaxy image. Middle - [\ion{O}{3}] image. Right: magnified MERLIN 5 GHz  radio image from \citet{akujor91}. Middle row: left - flux map for the broad velocity component, middle - velocity map of the broad velocity component, right - velocity dispersion map of the broad velocity component. Bottom row: same as middle row but for the narrow velocity component.}
\label{3C93_1map}
\end{figure*}

\begin{figure*}[t]
\centering
\includegraphics[width=6.in]{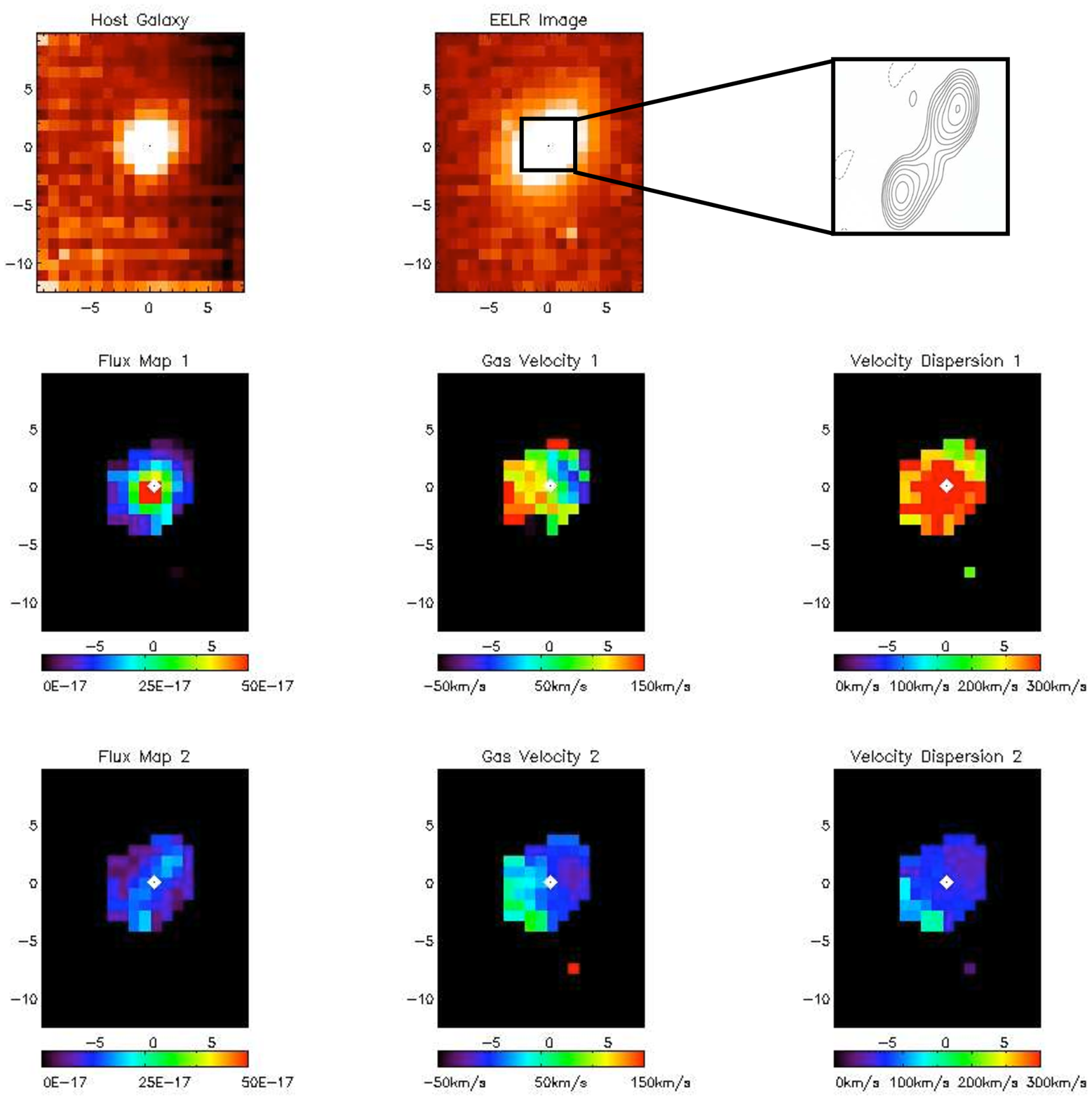}
\caption{{\bf PKS 0023-26} Top row: Left - host galaxy image. Middle - [\ion{O}{3}] image. Right: magnified MERLIN 5 GHz  radio image from \citet{tzioumis02}. Middle row: left - flux map for the broad velocity component, middle - velocity map of the broad velocity component, right - velocity dispersion map of the broad velocity component. Bottom row: same as middle row but for the narrow velocity component.}
\label{PKS0023map}
\end{figure*}

\begin{figure*}[t]
\centering
\includegraphics[width=6.in]{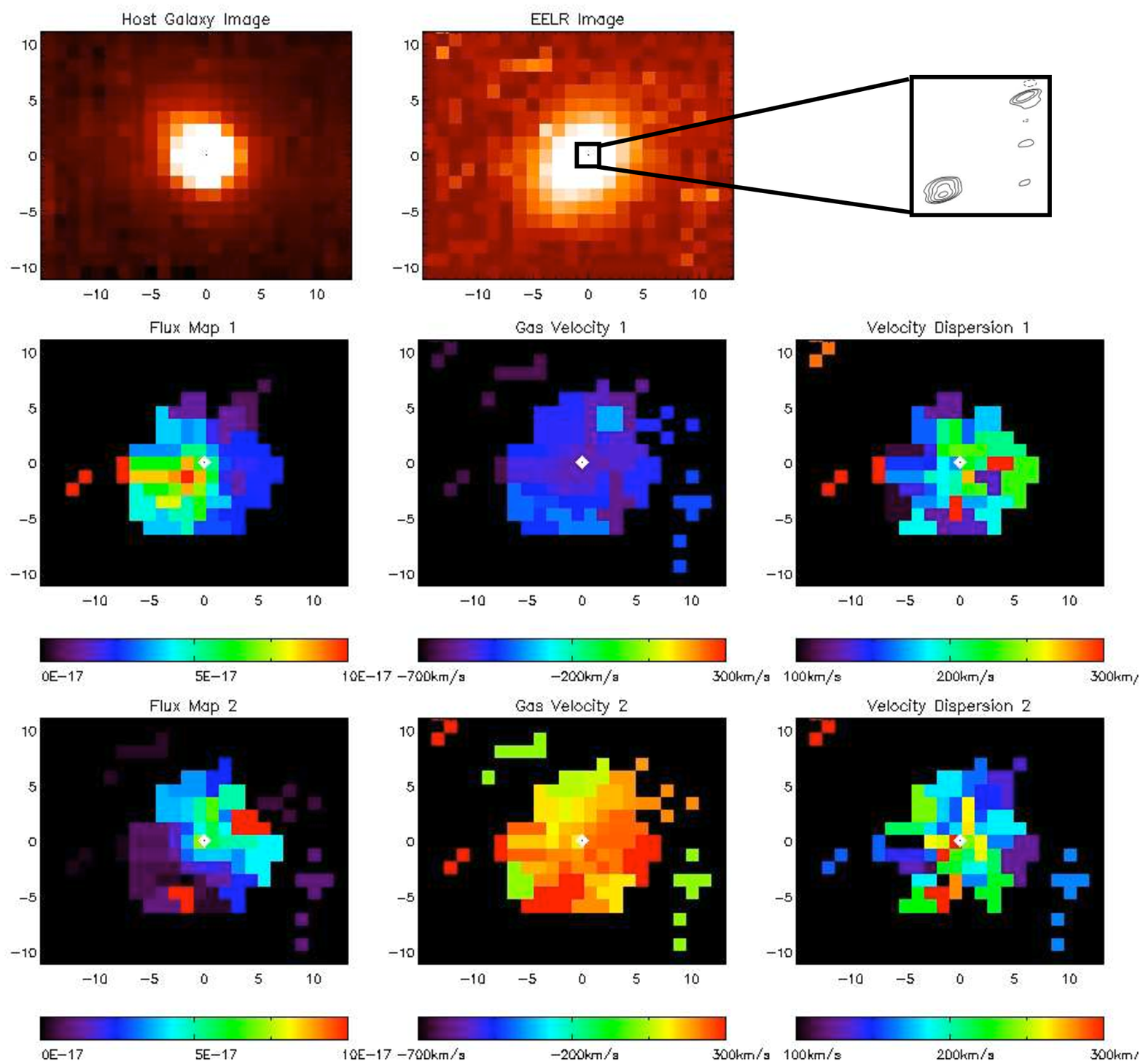}
\caption{{\bf PKS 1306$-$09} Top row: Left - host galaxy image. Middle - [\ion{O}{3}] image. Right: magnified MERLIN 2.291 GHz  radio image from \citet{tzioumis02}. Middle row: left - flux map for the broad velocity component, middle - velocity map of the broad velocity component, right - velocity dispersion map of the broad velocity component. Bottom row: same as middle row but for the second velocity component.}
\label{PKS1306-09map}
\end{figure*}

\begin{figure*}[t]
\centering
\includegraphics[width=6.in]{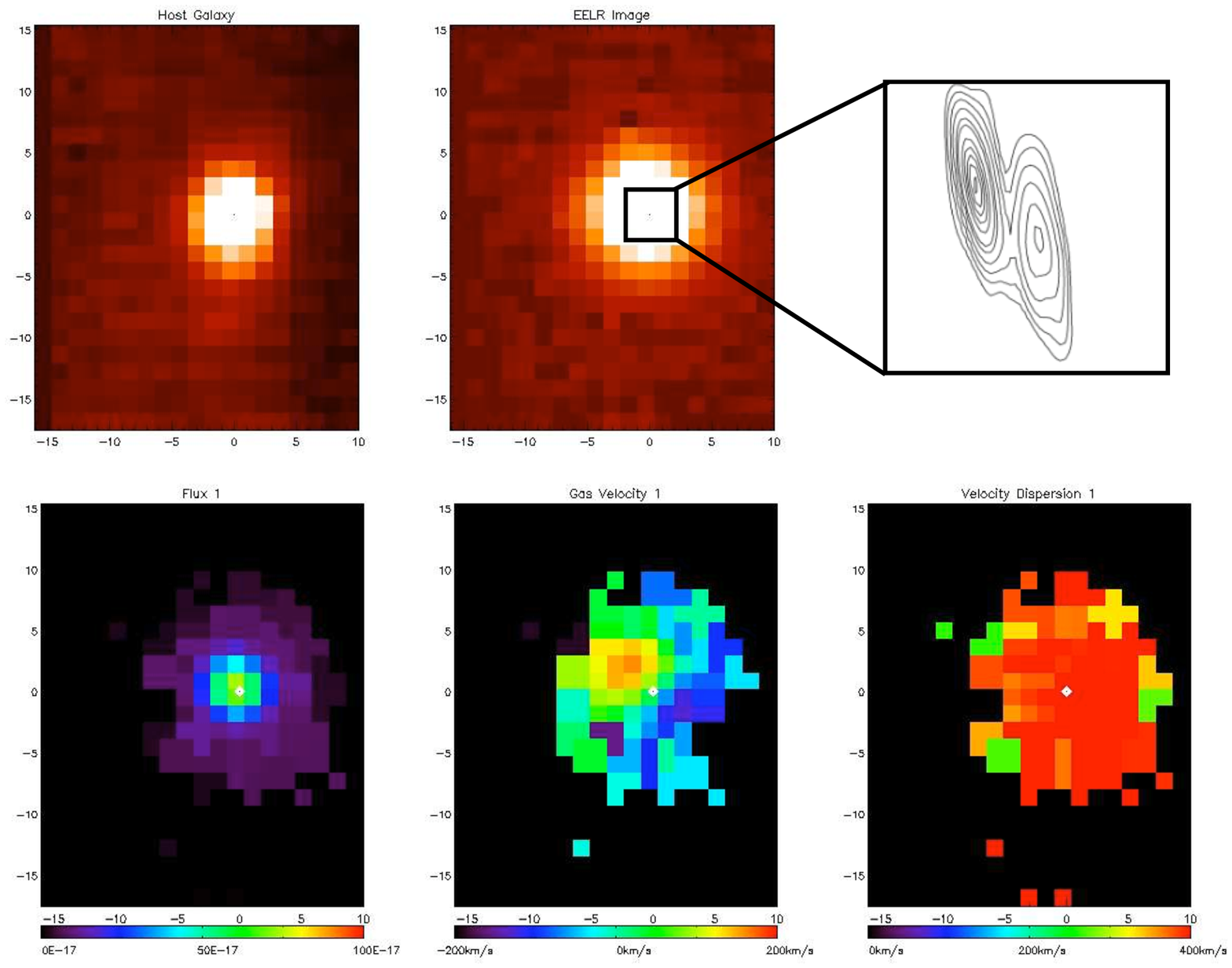}
\caption{{\bf PKS 2135$-$209} Top row: Left - host galaxy image. Middle - [\ion{O}{3}] image. Right: magnified MERLIN 5 GHz radio image from \citet{tzioumis02}. Bottom row: left - flux map for the broad velocity component, middle - velocity map of the broad velocity component, right - velocity dispersion map of the broad velocity component. }
\label{PKS2135-209map}
\end{figure*}

In this section we discuss the fit results for all the objects with detected EELRs. Figures \ref{3C49map1} to \ref{PKS2135-209map} display the EELR image, continuum image and emission-line fit results of each object. The EELR images are made by summing the continuum-subtracted slices on the [\ion{O}{3}] $\lambda 5007$ line. For each emission-line component, we show its flux distribution, velocity map and velocity dispersion map. The velocity dispersions are given in the standard deviation of the fitted Gaussian ($\sigma = ~$FWHM / 2$\sqrt{2\mathrm{ln}(2)}$), and they have been corrected for instrument resolution. 

The velocity maps show gas velocities with respect to the systemic velocity. We calculate the systemic velocity using the highest precision redshift we can find in literature. \citet{holt08} calculated high precision redshifts using all the narrow emission lines they could find in their nuclear spectra. We use the \citet{holt08} redshifts where our samples overlap. The other redshifts are drawn from sources listed in Table 1. These systemic velocities are either low precision or calculated from emission lines that may be contaminated by emission from outflowing gas. They can be used as rough reference points, but should not be interpreted as precise systemic velocities. Nevertheless, a conservative estimate suggests that they should be accurate to within $\pm 150$ km/s. 

The radio maps are shown either overlaid on the emission-line image or, in cases where it is too compact, enlarged beside the emission-line images. With no other stars in the IFU field, it is difficult to do a very precise overlay of the radio and optical image. Most sources with EELRs have two radio lobes and we assume that the mid-point of a line connecting the two lobes coincides with the peak of the continuum emission. For the more compact objects with no distinct lobes, we assume that the center of the radio structure coincides with the continuum peak. For this reason, we will only discuss the general alignment of the optical features to the radio axis. 

Most EELRs show a preferred direction of  elongation. We measure the PA of the elongation axis by fitting an ellipse to the contour connecting the pixels with roughly constant S/N. To determine the velocity gradient axis PA, we fit a plane of the form $z(x,y) = ax + by + c$ to the velocity field. A simple plane is only a rough approximation to the true geometry of the position-velocity diagrams, which sometimes involve additional curvatures. However, a visual inspection of the fits show that the value tan$^{-1}(b/a)$ suffice to provide the direction of maximum change in velocity per pixel. We apply an S/N cut of 10 the the pixels used for velocity gradient calculation. We report the measured velocity gradient PA in Table 2, along with the errors which take into account the uncertainties on line profile fitting and the effects of applying different S/N cuts.

{\bf 3C\,49} -This object has a complicated velocity field that requires up to four Gaussian components to fit. One component is clearly broader than the others, while the other three have smaller velocity dispersions of a similar range. Each component has a velocity range of a few hundred km/s, but there are no clear velocity gradients along any axis. The flux of the first three components shown in Figure \ref{3C49map2} peaks close to the center. The fourth has a peak to the west of the center, with some faint emission east of the center.  

The radio structure of 3C 49 consists of two hot spots oriented in the east-west direction spanning a total of 1.4\arcsec. The western lobe is bright, compact and roughly circular. The eastern lobe is fainter, more irregular, with a slight bend toward the north. The broad emission component is associated with the eastern lobe and so are two of the narrow components. The most redshifted narrow component is spatially the closest to the western lobe. The range of emission-line velocity components in the west may indicate that the 3-D morphology of the radio structure is complicated and involves more than one ``branch'' aligned at different angles to our line of sight. 

. 

{\bf 3C\,268.3} - There are two components to the emission line profile: one with a higher velocity dispersion and more concentrated toward the center of the source, and one with a lower velocity dispersion more spread out along the emission-line region. While the broader component has velocity fairly close to the systemic velocity, the narrower component has a clearly more blueshifted region and a clearly more redshifted one; there is a sharp transition between the two regions.  

The radio map of this object shows two asymmetrical lobes with PA of $\sim -20$\arcdeg. The EELR is elongated along a PA of $\sim -37$\arcdeg, slightly offset from the radio axis, but well aligned with velocity gradient axes of both components.

{\bf 3C\,303.1} - This object has two components, one broader and one narrower. The broader component is brighter toward the center, while the narrower one extends out in a roughly conical shape toward the north and south of the center. The radio axis is at PA of $-50$\arcdeg. The overall orientation axis of the EELR has a PA of $-32$\arcdeg, offset from the radio axis. The velocity axis gradient of both velocity components are well aligned with the radio axis.

{\bf 3C\,93.1} - This object's emission profile again consists of one narrow and one broad component. They are both concentrated close to the center, with the broad component brighter at the northeast and the narrow component brighter at the southwest. Both components have a velocity gradient at a PA of $\sim 21$\arcdeg, consistent with direction of the EELR's elongation. The radio structure of this object is complex and compact ($\sim$ 0.6\arcsec), with no apparent radio lobes or well defined radio axis. The optical emission-line region clearly extends beyonds the radio emission.

{\bf 3C\,124} - The EELR of this object has two distinct blobs, one in the north and one in the south. The emission lines require two components to fit, but both components have similar velocity dispersion. The more blueshifted component dominates in the north, while the more redshifted component dominates in the south. The radio emission of this object is elongated along the north-south axis. The two components can be interpreted as the two sides of an outflow. The flux weighted combined velocity map yields a velocity gradient with PA of $4$\arcdeg, well aligned with the radio axis. 

{\bf PKS\,1306$-$09} - The emission lines of this object require two components to fit. Both components have similar velocity dispersion. One component is clearly blueshifted, while the other is clearly redshifted. The blueshifted component is brightest at the southeast and the redshifted component is brightest at the northwest. The two components can be interpreted as the two sides of an outflow. 

The radio emission has two distinct components separated by 0.37\arcsec and has a PA of $-45$\arcdeg. The PA of the elongation axis of the EELR is $-60$\arcdeg. We measure the velocity gradient from a flux weighted combined velocity map of the two components. The velocity gradient axis PA is $-57$\arcdeg, consistent with the elongation axis PA.

{\bf PKS\,2135$-$209} - While most of the other objects have one broad and at least one narrow velocity component, no multiple components are resolved for this object. This is the only EELR object in our sample that requires only one component fit to its [\ion{O}{3}] line profile. The radio structure has two lobes at a PA of $55$\arcdeg. The PA of the velocity gradient axis is at 47\arcdeg, consistent with the radio axis within the error. The [\ion{O}{3}] emission appear circular, with no preferred direction of elongation.

{\bf PKS\,0023$-$26} - The emission lines of this object also have two distinct components, one broader and one narrower. The broader component is concentrated in a nearly circular shape around the center. The narrower component is elongated along an axis with PA of $\sim -45$\arcdeg. The radio emission from this object comprises two rather symmetric lobes aligned along an axis with PA of $\sim -26$\arcdeg, so the elongation axis of the EELR is slightly misaligned with the radio PA. The velocity gradient axes of both components are clearly offset from the radio PA and the elongation axis, with the narrow component almost aligned with the east-west axis.

\section{Discussion}

\subsubsection{Velocity Components}

The velocity structure of the EELRs are complex and can be separated into several different categories. Most objects have multiple velocity components. Some have distinct broad and narrow components, and others have two components of similar line width. There are also a few exceptions such as 3C\,49 and PKS\,2135$-$209. 

3C\,268.3, 3C\,303.1, 3C\,93.1, PKS\,0023-20 each have one broad and one narrow component that are spatially separated. The flux of the broad components tend to be more concentrated toward the center and the narrow components are more spread out. Except for PKS\,0023-20, the two velocity gradients for the same objects are aligned along the same axis. The slopes of the velocity gradient, however, are not necessarily the same. For example, 3C 268.3 has a narrow component with a velocity gradient of $143 \pm 8$ km s$^{-1}$ kpc$^{-1}$ close to the center where the velocity transition occurs, and the broad component has a gentler slope of ($46 \pm 10$ km s$^{-1}$ kpc$^{-1}$). The alignment of the velocity gradient axes suggests that the radio jets play a role in influencing the outflows for both components, but the flux distributions and the magnitude of the velocity gradients indicate possibly distinct origins. 

The EELRs of PKS\,1306$-$09 and 3C\,124 have two velocity components with similar velocity dispersions but a different velocity range. PKS\,1306$-$09 has one component at around $-500$ km s$^{-1}$, another at $+100$ km s$^{-1}$, both with velocity dispersions of $\sim 200$ km s$^{-1}$. 3C\,124 has velocity components centered at around $-700$ km s$^{-1}$ and  $+100$ km s$^{-1}$, both with velocity dispersions of  $\sim 350$ km s$^{-1}$. The flux distributions of the two components are concentrated opposite of each other along the radio axis. They can be interpreted as a two-sided outflow driven by the radio jet. 

Only one component is resolved for PKS\,2135$-$209. It is concentrated at the center and its velocity gradient is aligned with the radio axis. This can be a recently triggered outflow that did not yet have time to expand, or a two-sided outflow viewed as compact due to projection effects. However, since no bright radio core is detected in this object, it is unlikely that this object has jets aligned close to our line-of-sight. 

The most complex object, 3C\,49, has four components with different velocity ranges. Each component has distinct flux concentrations. This could be caused by a multi-branched radio morphology or strong interactions between the outflow and ISM gas.

\subsection{EELR Velocity Structure and Morphology}

The overall morphological alignment of EELRs with the radio axes in CSS sources has been shown by previous studies \citep[e.g.][]{privon08, axon00,devries99}, and confirmed in our data. In 3C\,303.1, 3C\,268.3, PKS\,0023-26, where the narrow components are resolvable, they are shown to be aligned with the elongation axis of the EELRs. We cannot resolve the broad components for most objects, except for 3C\,303.1, where it is shown to be also aligned with the elongation axis.

Although emission-line images of CSS sources have shown EELR gas that appears to extend more or less along the radio axis, it is difficult to determine the relationship between the gas and the radio jet without the full velocity map of the EELR gas. The gas can be an outflow driven by the jets, or gas dominated by gravitational motions (e.g. interstellar gas preferentially ionized by the AGN due to the radio jets clearing a path for the radiation, or gas accreted during recent galaxy interactions). The gas in the latter case would not have received significant energy or momentum transfer from the jet, and the velocity gradient of the gas would not show a strong correlation with the radio structure. 

\citet{holt08} have done an optical long slit study of CSS sources and determined velocities of the EELR gas along the slits. The objects that overlap with our sample are 3C\,303.1, 3C\,268.3, PKS\,0023-26, PKS\,1306-09, and PKS,2135-20. Where the sampling regions overlap, our emission-line profiles agree. With our the IFU data, it is possible to determine the velocity gradient axis for each velocity component. Three of our objects, 3C\,303.1, 3C\,268.3, and 3C\,93.1, each have one narrow and one broad component. Both velocity gradient axes are consistent with each other for all three objects. 

PKS\,1306-09 and 3C\,124 each have two components with similar velocity dispersions. We measure one velocity gradient for each object by using a combined, flux-weighted velocity map. The velocity gradient axes of both objects are closely aligned with the radio axes. 

The three remaining objects are 3C\,49, 3C\,93.1, and PKS\,0023-26. 3C\,49 has an unusually complicated velocity field and is not a typical object. 3C\,93.1 does not have a well defined radio axis, but both velocity axes gradients are well aligned with each other as well as the EELR elongation axis. They extend roughly perpendicular to the elongation of the host galaxy, which is at a PA of $\sim 135$\arcdeg ~\citep{devries97,devries99}. The velocity gradient axes for both components of PKS0023-20 are significantly offset from the radio axis and the elongation axis. The velocity gradient for this object is much smaller than for the others. If the extended emission is from a jet-driven outflow, it is likely that most of the gas motion is parallel to the plane of the sky and thus our velocity measurements are not the best indicators of the gas kinematics.

\subsection{Possible Scenarios}

There are several possible ways to interpret our data. Every scenario has its strengths and limitations, which will be discussed in the following sections.

\subsubsection{Radio Jet-Driven Outflow}

In systems with strong radio jets, there are at least two mechanisms that can drive an outflow. In the simulations of \citet{sutherland07}, the radio jets go through several phases when expanding. The emerging jet first finds its way through the disk by filling up the lower density regions in between denser gas clouds. As the jet injects energy into the surrounding medium, a high pressure, roughly spherical bubble forms and expands to larger than the scale of the disk. Most of the energy from the jet goes into feeding the bubble during this phase. The blast wave created by the expansion of the bubble can sweep up gas and become a driving force for a wide-solid-angle outflow. When the jets are about $55 - 70$ kyr old, they pierce through the bubble and start to form the classical radio lobes. The jets themselves may carry outflow material as they expand. 

For those systems with both narrow and broad components, one component can be driven by the collimated jets, and the other by the roughly spherical blast wave. Gas receiving energy and momentum transfer from the radio jets have velocity vectors pointing predominately in the direction of the jet expansion. Therefore, unless the jets are pointing directly along our line of sight, the projected velocity dispersion we observe will be small relative to a spherically expanding gas cloud. 

The outflow generated by the blast wave should have a much wider opening angle. As in the case of 3C\,48, the broader component's line profile may indicate a wide-opening-angle outflow where the projected velocity vectors encompass a large range of wavelength shifts. While a radio jet can continuously entrain the gas that it encounters as it expands, a blast wave is a more centrally concentrated energy source. For those objects without a distinct broad component, such as PKS 1306$-$09, the blast-wave driven outflow is not necessarily absent, but it may not be detectable in our data due to a smaller amount of gas available at the center. 

The majority of the velocity gradient axes measured are closely aligned with the radio jets, which further strengthens the argument where the kinematics of the EELRs are dominated by influences from the radio jets. 3C\,303.1 and 3C\,268.3 are identified in \citet{holt08} to have broad profile, high velocity outflows, which correspond to the centrally concentrated broad components in our data. \citeauthor{holt08} suggests that the narrow emissions are from ambient, quiescent ISM. While that remains a possibility, the alignments of velocity gradient axes in our data suggest that it is more likely for both broad and narrow velocity components to be receiving energy and momentum transfer from the same source. Note that although this is a reasonable starting point given the trends seen in our data, this is a very simplified picture for a rather complicated system and it may not be able to explain everything observed in the data.

The slight misalignments between the EELRs and the radio structures may be explained by jet interaction with inhomogeneous ISM gas. \citet{wagner11} have shown through a simulation that the progression of jets through a dense clumpy ISM is very different from those encountering a homogeneous ISM. In a homogeneous environment, the jet pierces through the cloud following a straight and narrow path. When the surrounding gas clouds are porous and have dense clumps, the jet gets deflected throughout the cloud, finding the path of least resistance while gradually dispersing the dense clouds. The cloud material can be accelerated up to 1000 km s$^{-1}$. In the case of clumpy ISM, the gas driven outward by the jets extends out much farther from the jet axis and can be traveling in a wider range of directions. In some objects, it is possible that the inhomogeneous ISM causes the radio jet and/or the outflowing gas to lose their alignment. However, this type of cloud-gas interaction will more likely broaden the cone of the outflowing gas instead of completely redirecting the outflow. We currently do not know of a good scenario where outflowing entrained gas can decouple from the radio jets. On the other hand, an outflow channeling through a clumpy medium offers a plausible explanation for the multiple branches of EELR gas in 3C 49. 

Other than the misalignments, there are also a few cases where the EELRs extend beyond the radio structure (e.g. 3C\,268.3). It is unlikely that outflows driven by the radio jets can extend further than the jet themselves. This indicates that other mechanisms may be at work.

\subsubsection{Multiple Episodes of Jet-Driven Outflow}

If there are multiple episodes of radio activity, the extended gas beyond the radio structure may be outflows driven by the previous set of radio jets. 3C\,48 is an example where there might be outflows driven by multiple episodes of radio activity. It is a CSS quasar with high velocity outflow at the base of its radio jet. It also has extended emission far beyond the radio structure north of the quasar, consistent with the direction the radio jet is pointing \citep{stockton87, axon00}.

If the radio jets from both episodes are pointed in the same direction, this scenario can explain the alignments of velocity gradient axes even for emission-line gas beyond the radio structure, assuming that the UV radiation from the central source can escape in the same general direction of the jet. However, the EELRs in our data appear to be smooth and continuous, instead of the more isolated clumps of extended gas expected from different episodes of radio activity.

\subsubsection{Photoionized ISM and Merging Tidal Tails}

There is also the possibility that the EELRs consist, partially or entirely, of non-outflow gas. The apparent alignments of the EELRs' elongation axes and the radio axes may be due to the radio jets clearing out a path for the AGN radiation to preferentially ionize gas that are close to the jet's path. The gas can be in situ, or accreted in recent merging events. This scenario can easily explain the narrow emission-line regions extending beyond the radio structures. However, the velocity gradient axes the the radio axes can only be aligned by pure chance. 

Most velocity gradients observed are closely aligned with the corresponding radio axes within $15$\arcdeg. The probability of such alignment occurring by chance for one galaxy is $\sim26\%$. We have five galaxies, 3C\,268.3, 3C\,303.1, 3C\,124, PKS\,1306-09,  and PKS,2135-20, with such alignments. If we do not count 3C\,93.1 due to its lack of a clear radio axis, five out of seven of our targets have velocity gradient axes closely aligned with the radio axes. The probability of finding this by chance is $\sim 5\%$.

\subsection{Emission Line Ratios}

\subsubsection{AGN Photoionization}
Given our limited wavelength coverage, complete line diagnostics are not possible for most of the objects. We map the line ratios for six objects that have good S/N ($> 50$ at the brightest spaxel) for both  [\ion{O}{2}] $\lambda 3727$ and [\ion{O}{3}] $\lambda 5007$ lines. Figure \ref{O3O2_shockmodels} shows the [\ion{O}{2}]/[\ion{O}{3}] ratios for each object. Differential atmospheric refraction causes the image position on the data cubes to shift as a function of wavelength. To correct for the shift, for each line, we take the median continuum immediately to the red and blue side of the line to determine the continuum centroid. The median continuum images are also used to scale and subtract the continuum contribution to the emission-line images. The ratio image is masked such that only the regions with bright emission lines are shown, filtering out the low S/N region of the data cube. The [\ion{O}{2}]/[\ion{O}{3}] ratios are indicators of ionizations state; lower ratios will give higher ionization parameters. Measurement of reddening effects is only possible in 3C\,303.1 via the H$\gamma$/H$\beta$ ratio, which is measured to be $0.51 \pm 0.14$, consistent with no reddening. All following discussions related to the emission-line ratios assume no reddening.

To constrain the ionization parameter and abundances, we plot the line ratios on a diagnostic diagram along with dust-free AGN photoionization model grids from \citet{groves04}. In these model grids, the spectral energy distribution (SED) of the ionizing source is represented by simple power laws with spectral indices of $\alpha$ = $-1.2, -1.4, -1.7$ and $-2.0$. The total radiative flux is characterized by the ionization parameter $U_{0}= S_{*}/(n_{0}c)$, where $S_{*}$ is the total flux of ionizing photons entering the clouds, $n_{0}$ is the initial density of the clouds and $c$ is the speed of light. The log ($U_{0}$) values included in the model grids ranges from $-4.0$ to 0.0.

Figure \ref{O3O2_shockmodels} shows the [\ion{O}{2}]/[\ion{O}{3}] vs. [\ion{O}{3}]/H$\beta$ plot for all objects. The black model grid is the dust-free AGN photoionization with $Z=0.5$Z$_{\odot}$ and $n=100$. The error for each emission line is calculated based on the photon counts; it is propagated to the emission-line ratio plot.  Points with low signal-to-noise, and therefore large error bars, are removed from each plot. The results show that the high S/N data points form tight clumps on the line ratio diagram, spanning different regions for each object. For example, PKS 1306$-$09 clearly lands in a region with lower ionization parameter than the other objects. For 3C\,303.1, 3C\,93.1 and PKS\,2135-20, the AGN photoionization grid do not encompass all the data points. Previous studies suggest that the EELR gas around CSS objects may be ionized by a mixture of shocks and AGN radiation \citep[e.g.][]{labiano05, holt09}. We will discuss this possibility of shock ionization in the next section.

\subsubsection{Shocks}
\begin{figure*}[t]
\centering
\includegraphics[width=6.in]{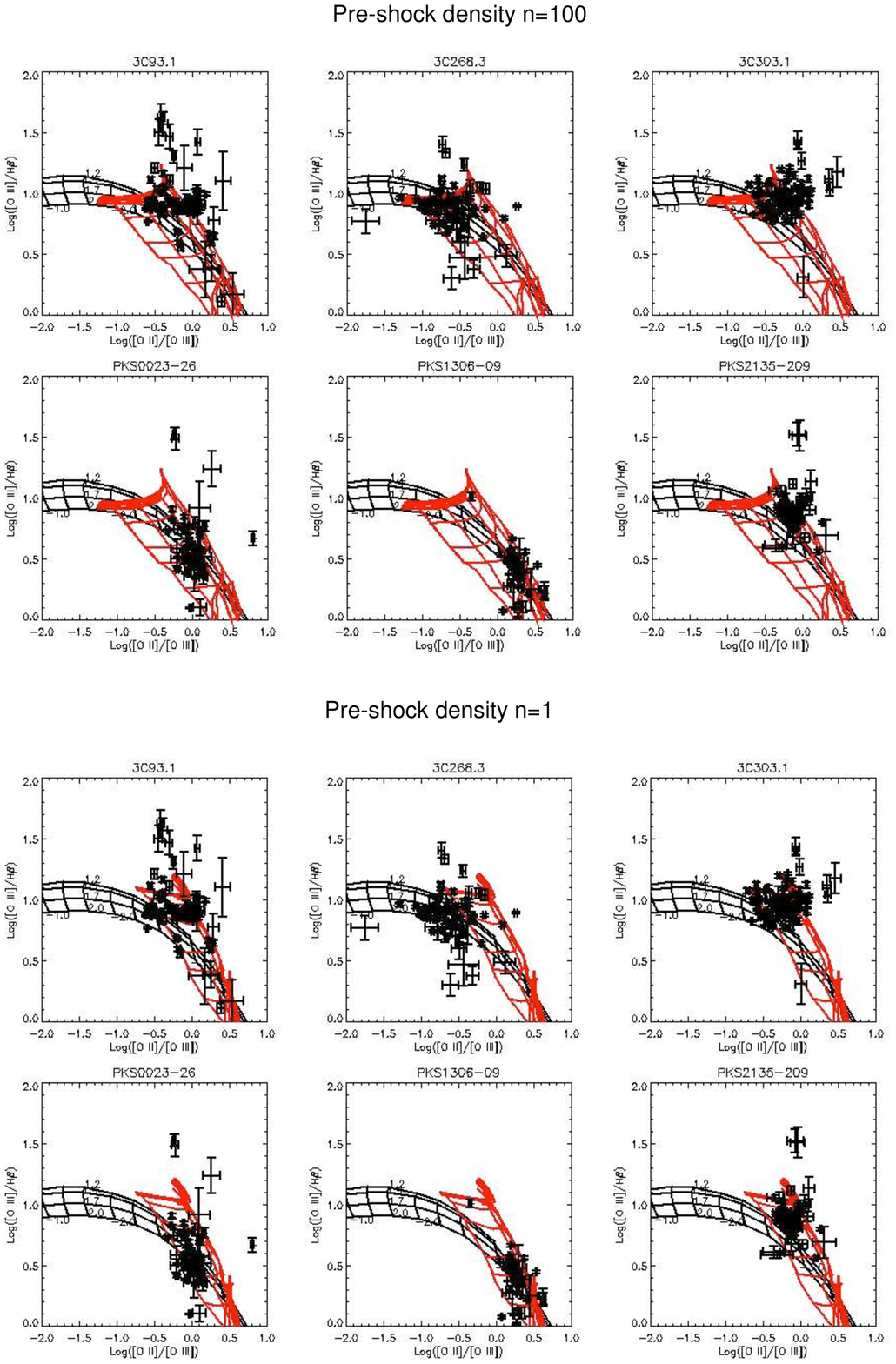}
\caption{Top: Shock + precursor (red) models with pre-shock density of $n=1$ and solar metallicity and dust free AGN (black) with density of $n=100$ and $Z=0.5$Z$_{\odot}$.  Bottom: Same as top except pre-shock density is n$=100$ for the shock model. }
\label{O3O2_shockmodels}
\end{figure*}

In EELRs around more extended radio sources, shocks have been found to be an insignificant source of ionization relative to AGN photoionization. The UV spectrum of 4C\,37.43 obtained by \citet{stockton02} shows that the \ion{O}{4} $\lambda1402$ and \ion{C}{4} $\lambda1549$ lines are too weak relative to \ion{He}{2} $\lambda 1640$ for shock ionization to be significant. However, at the early stages of the radio jet evolution the effects of shocks may be more important. Expanding high pressure bubbles compress the gas clouds, which result in shocks. The jets themselves can also cause shocks when they encounter dense clumps in the confining medium, which cool through shock-excited emissions \citep{sutherland07}. Models of jet feedback often include non-thermal-equilibrium cooling as a result of shocks \citep{wagner11}. We investigate the possibility of shocks by plotting shock model grids from \citet{allen08} on our data points. In Figure \ref{O3O2_shockmodels}(a) a significant fraction of the data points lie well outside of the AGN ionization model grids. The overlaid shock + precursor models \citep{allen08} encompass a larger fraction of the spaxels ($\sim 70\%$ in the case of 3C\,303.1). Decreasing the pre-shock density to $n=1$ cm$^{-3}$, shown in Figure \ref{O3O2_shockmodels}(b), appears to encompass more spaxels for 3C\,303.1 ($\sim 90\%$ of the spaxels), 3C\,93.1, and PKS\,2135$-$209.

Even though shock models appear to give a better overall fit to the data, it is difficult to disentangle the effects of shock ionization from AGN photoionization with the data that we have. Further line diagnostics, with more line ratios such as [\ion{N}{2}]/H$\alpha$ or UV lines, are required. However, given the physical locations of the EELR and the stage of evolution the CSS objects are at, it is possible that both mechanisms are significant. 

\begin{figure*}[t]
\centering
\includegraphics[width=6.in]{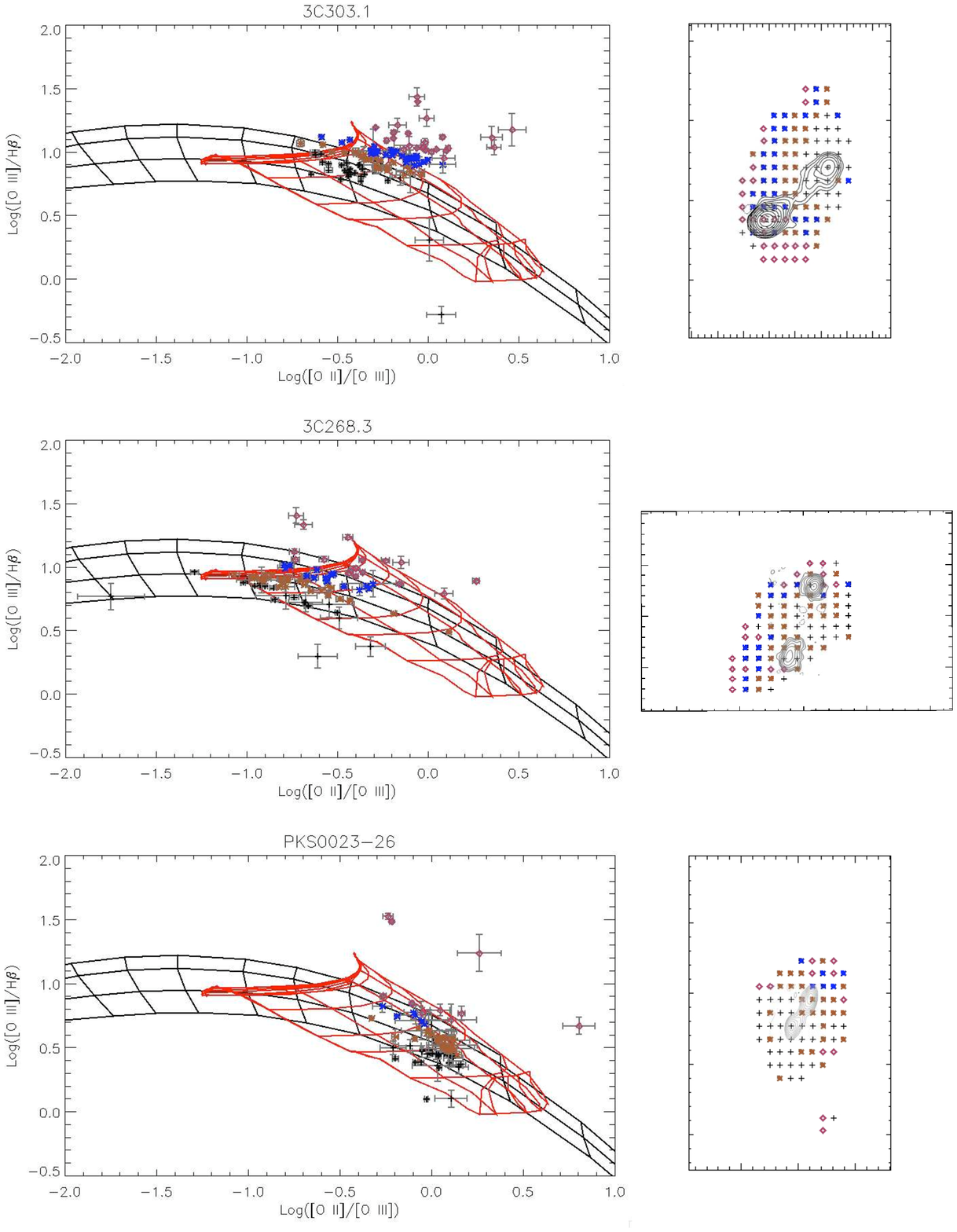}
\caption{Top row - left: The line ratio plot for 3C 303.1 showing the data points divided by distance from the AGN photoionization grid. Both model grids have solar metallicity and $n=100$. Right: The spatial plot of the divided data points. Middle and bottom row - Same as top row but for 3C 268.3 and PKS 0023$-$26 respectively. }
\label{line_radio1}
\end{figure*}

To visualize a possible interplay between AGN and shock ionization, we divide the data points in the [\ion{O}{2}]/[\ion{O}{3}] vs. [\ion{O}{3}]/H$\beta$ for all objects into 4 sections with increasing distance from the lowest spectral index AGN photoionization grid and plot them in spatial coordinates. The diagnostics for 3C\,303.1, 3C\,268.3 and PKS\,0023$-$26 are shown in Figure \ref{line_radio1}(a), (b) and (c) respectively. The radio contours of the objects are over-plotted on the spatial maps. Other objects do not show as clean spatial patterns when the data points are divided.

{\bf 3C\,303.1} - We see a spatial pattern with this type of division: The points that fall inside the AGN grid concentrate in a northwest region. With increasing distance from the AGN grid, the data points also approach lower ionization parameters and the $\sim 300$ km s$^{-1}$ region of the shock grid. One interpretation of this pattern, since the gradient appears to correspond to the radio structure, is that the northwest radio lobe cleared out a path for AGN radiation which dominates the gas ionization in the northwest region. The other regions have less AGN radiation penetrating through and are increasingly dominated by shock ionization. 

{\bf 3C\,268.3} - Almost all data points for this object fall into both model grids. However, with increasing [\ion{O}{3}]/H$\beta$ ratio, the data points also shift not only toward lower ionization parameters, but also lower spectral index in the AGN grid. The spatial pattern shows that points with higher [\ion{O}{3}]/H$\beta$ tend to concentrate further from the continuum peak in the spatial plot. Since it seems unlikely that the spectral index of the AGN ionizing spectrum should change significantly in one object, the more likely explanation is that both shocks and AGN ionization are involved.

{\bf PKS\,0023$-$26} - While almost all data points fall inside of both model grids, there is again a clear spatial pattern to the divided data points. If the ionization in this object is all shock dominated, then this pattern indicates a gradient in shock velocity. It is unlikely for AGN ionization to be the only dominant ionization mechanism here since it would again require a range of spectral indices.

\subsection{Comparison with Classical EELRs}

Our analysis finds that, with a few exceptions, the EELR morphologies and velocity gradients are well correlated with the radio emission. Where samples overlap, our results are consistent with the HST imaging of EELRs around CSS sources in \citet{axon00} and \citet{privon08}. The classical EELRs around more extended radio sources, however, have been shown to have globally chaotic morphology and velocity structures \citep[e.g.,][]{stockton87, fu06, fu08} with no clear evidence for jet-cloud interactions. \citet{privon08} took emission-line images of 80 radio sources (including 4 CSS sources) and showed that while the difference in PA between the EELRs and the radio axes can span a wide range for extended FR II sources, this quantity tends to be $< 20\arcdeg$ for CSS objects. We have confirmed the close alignment for CSS objects with a larger sample size. The higher spread of relative alignments in more extended radio sources suggest a evolutionary trend where the EELRs evolve to become more disorderly and less correlated with the radio structure with time. 3C 49, with it's more chaotic velocity structure, may represent a transition object at a later stage of evolution than the rest of our sample. 

In classical EELRs, there are no indications of shock photoionization. In the study of \citet{fu06}, even though the overlapping model grids make distinguishing between AGN ionization and shock ionization difficult, the data can still be entirely explained by AGN ionization. In our observations, while there are still no clear shock signatures, it is not possible to fit all the data points with pure AGN ionization models, and adding a shock component results in a better fit. Previous work has also considered multiphase AGN photoionization models involving two media with different density and filling factors \cite[e.g.,][]{fu06,fu07b}. However, since shocks are very likely to be a significant source of ionization in a CSS source, a pure AGN ionization model may not be as physically plausible. Considering multiphase gas with both AGN and shock ionization has the potential of providing better fits, but it comes with additional free parameters that are unnecessary for explaining the current data that we have.

\section{Conclusion}

We have analyzed the velocity structure and emission-line ratios for all of the EELRs detected in our sample. The alignment of the EELR velocity gradient axes with the radio axes supports a scenario where the EELRs consist of outflowing gas driven by the radio jet-related activities. Emission-line gas extending beyond the radio structure may also contain contributions from previous episodes of radio activity, ionized ISM, and recently accreted gas. We find that for most cases, the EELRs have at least two velocity components with distinct velocity dispersions, velocity gradients and/or flux distributions. The distinct components may be outflows with different driving mechanisms, such as a pressure-driven blast wave vs. collimated radio jets. In some cases this scenario may be complicated by the interaction of the radio jet with the inhomogeneous gas clumps in the host galaxy disks. 

We have compared our emission line ratios with various photoionization models. Because of the proximity of the EELR gas to the AGN, and the various shock mechanisms related to the onset of the radio jets, the most likely scenario is that the extended gas is ionized by a combination of AGN radiation and shocks. Some objects display clear spatial pattern that can be interpreted as increasing influence by one ionization mechanism from one region to another. Other objects that do not display such pattern may have one dominant ionization mechanism throughout the whole EELR. More line ratios are required to further disentangle the effects between different ionization mechanisms and metallicities. 

This research is supported by NSF grant AST-0807900. This paper is based on observations obtained at the Gemini Observatory, which is operated by the Association of Universities for Research in Astronomy, Inc., under a cooperative agreement with the NSF on behalf of the Gemini partnership: the National Science Foundation (United States), the National Research Council (Canada), CONICYT (Chile), the Australian Research  Council (Australia), MinistŽrio da Cincia, Tecnologia e Inova‹o (Brazil) and Ministerio de Ciencia, Tecnolog'a e Innovaci—n Productiva (Argentina). L. Kewley acknowledges support from Australian Research Council Discovery Project DP 130104879 and NSF Early Career Award AST07-48559. We thank the anonymous referee for detailed comments that helped us improve both the content and the presentation of this paper.

\end{document}